\documentclass[proceedings]{JHEP3}

\conference{International Europhysics Conference on HEP} % Dont change this one

\usepackage{epsfig}

\newcommand{\gapprox}{\stackrel{>}{_{\sim}}}

\newcommand{\ra}{\rightarrow}

\newcommand{\bbb}{b\bar{b}}
\newcommand{\pbp}{\bar{p}p}
\newcommand{\epem}{e^+e^-}
\newcommand{\gaga}{\gamma\gamma}

\newcommand{\pbmo}{pb$^{-1}$}

\title{Heavy Quark Production in Deep-Inelastic Scattering}

\author{\speaker{Felix Sefkow}\thanks{On behalf of the H1 and ZEUS Collaborations}  \\
Physik-Institut der Universit\"at Z\"urich,
Winterthurerstr.\ 190, 8057 Z\"urich, Switzerland 
E-mail: \email{felix.sefkow@desy.de}
}                    

\abstract{We review recent results from the H1 and ZEUS experiments 
at HERA on charm and beauty production in $ep$ collisions 
at 300 - 318 GeV centre-of-mass energy.}

\begin{document}

  \section{Introduction}

Deep inelastic scattering (DIS) at HERA 
offers unique opportunities to test and refine our understanding
of heavy quark production in terms of perturbative QCD.  
The dominant mechanism 
here is boson-gluon fusion (BGF): a photon coupling to the scattered positron
interacts with a gluon from the proton to form a quark-antiquark 
pair. 
A quantitative description of this process requires  
the gluon momentum distribution in the proton, 
a partonic matrix element and a fragmentation function. 
The gluon density is known to an accuracy of a few percent from the analyses of 
scaling violations of the proton structure function $F_2$ 
measured at HERA~\cite{f2h1zeus}.
The masses of the charm and, even more so, of the beauty quark 
ensure that a hard scale is present 
that renders QCD perturbation theory to be applicable 
to the calculation of the hard subprocess.
Fragmentation functions,  which account 
for the long-range effects binding the heavy quarks in 
observable hadrons, are extracted
from $\epem$ annihilation data, 
where the kinematics of the hard process is well determined;
results with high precision appeared recently~\cite{alephsldbfrag}. 
Compared with the clean $\epem$ case, 
the complication in $ep$ collisions lies in the strongly interacting 
initial state. 
However, relative to other production environments like  
hadron-hadron collisions or two-photon interactions, uncertainties
related to hadronic structure are reduced to a minimum.  

QCD calculations have been performed up to fixed order $\alpha_s^2$
in the so-called massive scheme, where only gluons and light quarks are 
active partons in the initial state. 
They are available in the form of 
Monte Carlo integration programs~\cite{hvqdis}, which, by using 
Peterson fragmentation functions~\cite{peterson}, 
provide differential hadronic cross sections.
Due to the higher quark mass, the QCD predictions are expected to 
be more reliable for beauty than for charm. 
However, we note that the NLO corrections to the predicted
DIS cross section are around 40\% of the LO result in both cases. 
At very high momentum transfers, a treatment in terms of heavy quark  
densities in the proton may be more adequate;
but differences between these schemes are not yet significant in 
the range covered by HERA so far~\cite{smithvfns}. 

\EPSFIGURE{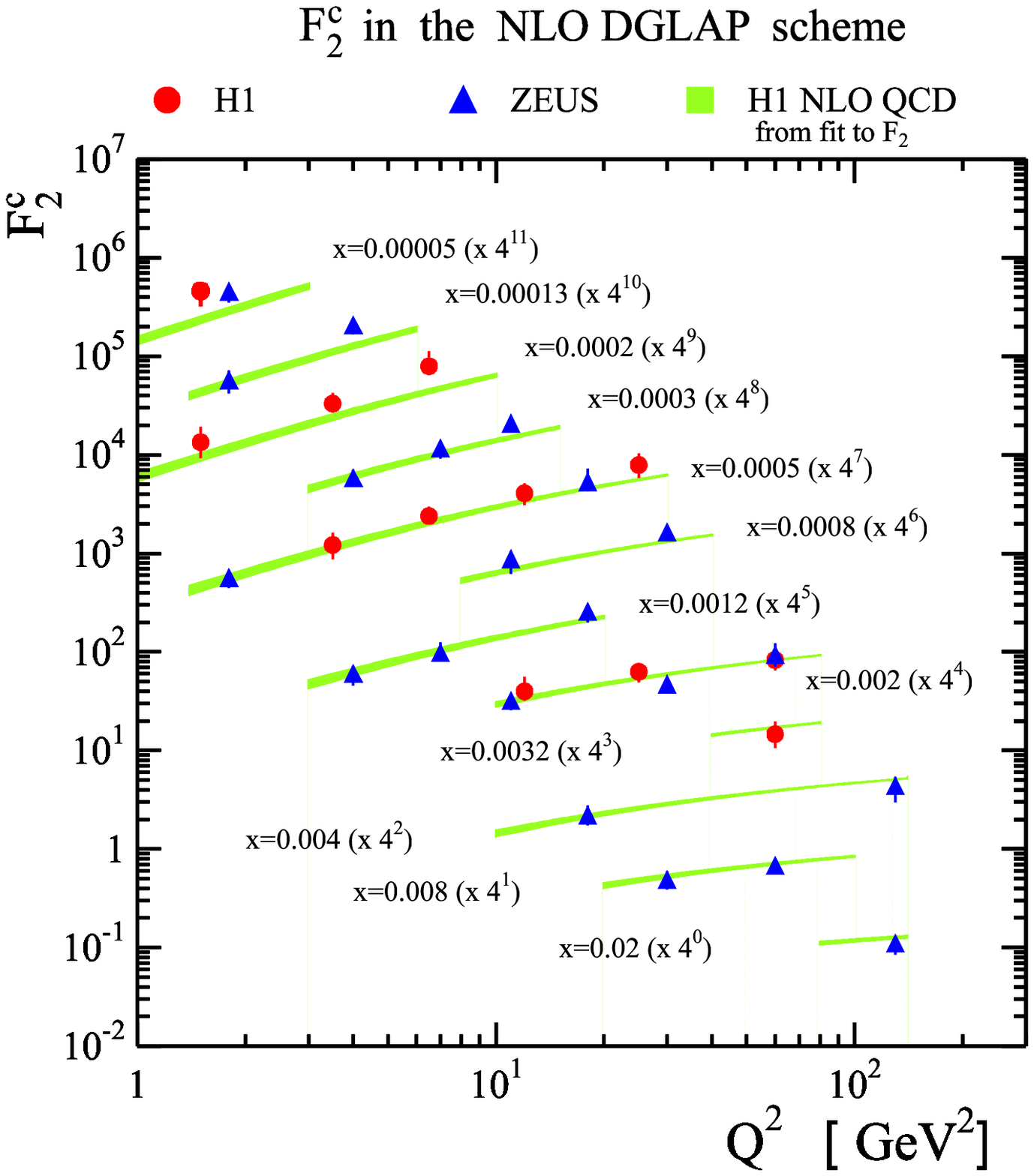,width=7.2cm}%
{\label{fig:f2scalv}
Charm contribution to the proton structure function, 
compared with NLO QCD.}
\section{Charm}
Most of the HERA results on charm make use of 
the ``golden'' decay channel $D^{\ast +} \ra D^0\pi^+ $ followed
by $D^0\ra K^-\pi ^+$; ZEUS also uses semileptonic decays.
The contribution of charmed final states to DIS 
is quantified as the ratio $F_2^c/F_2$, 
where $F_2^c$ is defined in 
an analogous way to the proton structure function $F_2$ by
\[\frac{d^2\sigma(ep\ra cX)}{dx\, dQ^2}
= \frac{2\pi\alpha^2}{xQ^4}
(1+(1-y)^2)\cdot F_2^c (x,Q^2)\; ;\]
$x$, $y$, and $Q^2$ are the standard DIS scaling variables. 
Measurements of $F_2^c$ 
by both experiments~\cite{zeusf2cdl,h1f2c} 
and by using different channels yield consistent results.  
The charm contribution $F_2^c/F_2$ is found to be about 20 to 30~\% in 
most of the kinematic region at HERA.
It is large where gluon-induced reactions dominate,
and decreases only as $Q^2$ becomes smaller than $\sim$10~GeV$^2$, or at higher
$x$ values $\gapprox 0.01$, 
where the quark content in the proton takes over.  
Figure~\ref{fig:f2scalv} displays the measured values of $F_2^c$.
The $Q^2$ dependence, for fixed values of $x$, is steeper 
than for the inclusive structure function.
% in fact charm accounts for about half of the scaling violations of $F_2$. 
The NLO QCD calculation~\cite{hvqdis}, 
with a gluon distribution extracted from H1 $F_2$ data,
agrees well with the data, 
which demonstrates the overall consistency of 
the boson-gluon-fusion picture. At low $x$, the data tend to be somewhat 
higher and to vary stronger with $Q^2$ than the prediction.  

\FIGURE{\unitlength1cm
\begin{picture}(7.,2.7)
 \put(0,-0.7){\epsfig{file=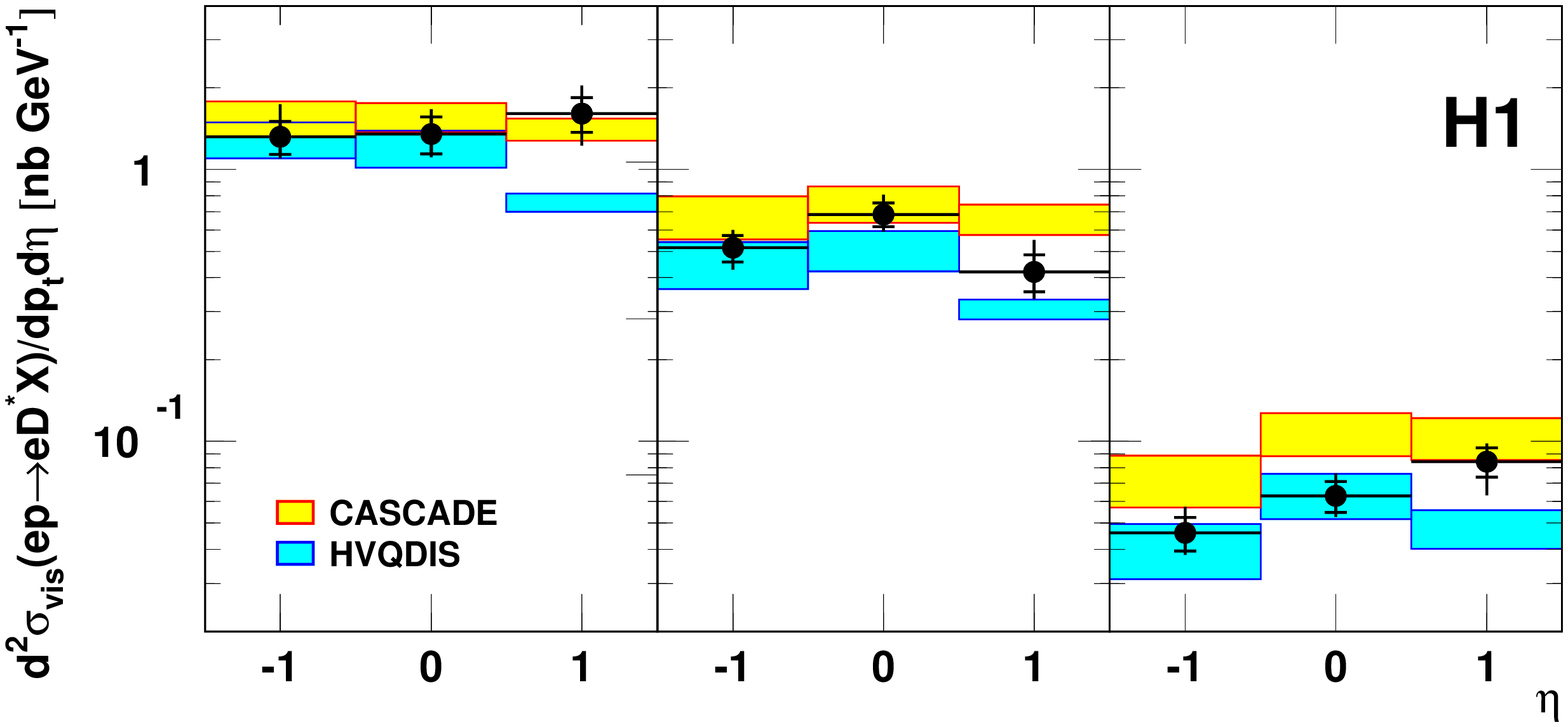,width=7.2cm}}
\put(1,2.9){\tiny\sf {$1.5<p_T<2.5$}}
\put(3.1,2.9){\tiny\sf {$2.5<p_T<4$}}
\put(5.1,2.9){\tiny\sf {$4<p_T<10\,$GeV}}
\end{picture}
\caption{\label{fig:dstddiff}
Double differential $D^{\ast\pm}$ cross section, % $d^2\sigma/d\eta\,dp_T$
compared with QCD predictions.}}
The available statistics makes more detailed investigations possible.
It was observed earlier that the NLO QCD calculations with Peterson fragmentation
(HVQDIS) do not reproduce 
the rapidity ($\eta$) 
distribution of the produced charm meson
well in the forward region 
(the outgoing proton direction)~\cite{zeusf2cdl,h1glue}.
The double differential $D^{\ast}$ cross section~\cite{h1f2c} displayed 
in Figure~\ref{fig:dstddiff} reveals 
that in the H1 data this is predominantly a feature of the low $p_T(D^{\ast})$ region.
%at larger $p_T$, 
%the HVQDIS calculation~\cite{hvqdis}, here using the 
%GRV parton distribution set~\cite{grv}, reproduces the data reasonably well. 
The measurement is also compared with the CASCADE Monte Carlo program~\cite{cascade}, 
based on the CCFM evolution equation~\cite{ccfm}, 
which resums higher order contributions at low $x$. 
Using an unintegrated gluon distribution extracted from inclusive H1 data,
it reproduces the data in the low $p_T$ region well, 
but overshoots at higher $p_T$. 
%Since the extraction of $F_2^c$ involves an extrapolation 
%to the full phase space, 
Such shape differences %in the predicted final state distributions  
imply that the extrapolated result for $F_2^c$ is model-dependent,
but also the $F_2^c$ prediction depends on the evolution scheme.    
H1 has performed a consistent extraction and comparison in the 
CCFM scheme and found somewhat better agreement in the low $x$ region 
than in the standard Altarelli-Parisi scheme.  

\FIGURE{\unitlength1cm
\begin{picture}(6.,3.4)
% \put(0,3.8){\epsfig{file=virt_gam.sas.eps,clip=,width=6.5cm}}
 \put(0,-0.5){\epsfig{file=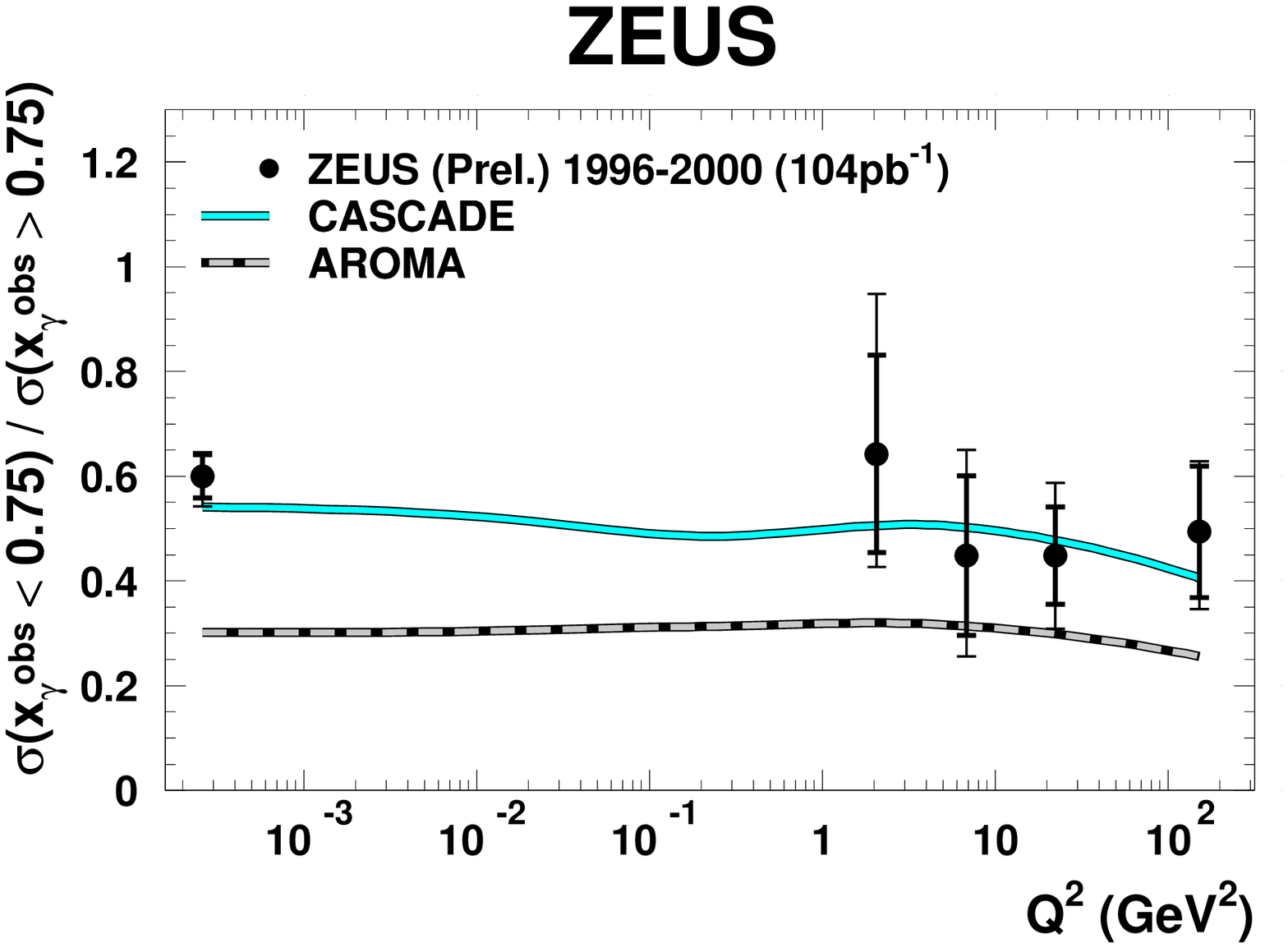,width=6.4cm}}
\end{picture}
\caption{\label{fig:virgam}
Ratio of low $x_{\gamma}$ to high $x_{\gamma}$ cross section,
compared with % Monte Carlo 
predictions.
}}
A possibility  to include higher order processes in the modeling of 
heavy quark 
production is to use the concept of photon structure also at non-zero 
virtuality.
One can classify
events as ``direct'' or ``resolved'' according to the measured value 
of $x_{\gamma}^{OBS}=(E-p_z)_{2\; jets} / (E-p_z)_{all\; hadrons}$
in dijet events. 
Keeping the LO photoproduction language, this corresponds
to the momentum fraction of the incoming parton in the photon, 
but  more generally, $x_{\gamma}^{OBS}$ is sensitive to 
any kind of non-collinear radiation in the event. 
The ratio of resolved {\it vs.}\ direct cross sections has been determined 
in this approach by ZEUS~\cite{zeusvirgam} and is displayed as a function 
of virtuality in Figure~\ref{fig:virgam}. 
In contrast to the situation for light quarks, 
the ratio in the DIS regime is very similar to that at $Q^2\approx 0$;
as expected, using e.g.\ the 
virtual photon structure function set SaS1D~\cite{sas}
implemented in the HERWIG program~\cite{herwig}.
The CASCADE model, 
with gluon emissions ordered in angle rather than in $k_T$, 
effectively incorporates the perturbative, 
anomalous part of the photon structure and reproduces the data well at all $Q^2$, 
but not the AROMA Monte Carlo program~\cite{aroma}, which does not include 
such contributions. 
%
%At momentum transfers much larger than the charm mass, 
%the fixed order calculation of charm production may become unreliable, 
%and a treatment in terms of charm parton densities may be more adequate. 
%ZEUS has measured the cross section up $Q^2=1000$ GeV$^2$ and $x=0.01$, 
%and found  that the boson gluon fusion picture 
%still provides a good description. 
%So-called variable flavour schemes, interpolating between the 
%two approaches, have been developed, but only at even higher $x$ and $Q^2$ 
%data will have sensitivity to different implementations~\cite{smithvfns}.
%
%\EPSFIGURE{dis_anom.q2.eps,clip=,width=6.cm}%
%{\label{fig:eminus}
%$D^{\ast\pm}$ cross section in $e^-p$ and $e^+p$ DIS
%{\it vs.}\ NLO QCD prediction.}
\FIGURE{\unitlength1cm
\begin{picture}(5.8,6.4)
 \put(0,-0.9){\epsfig{file=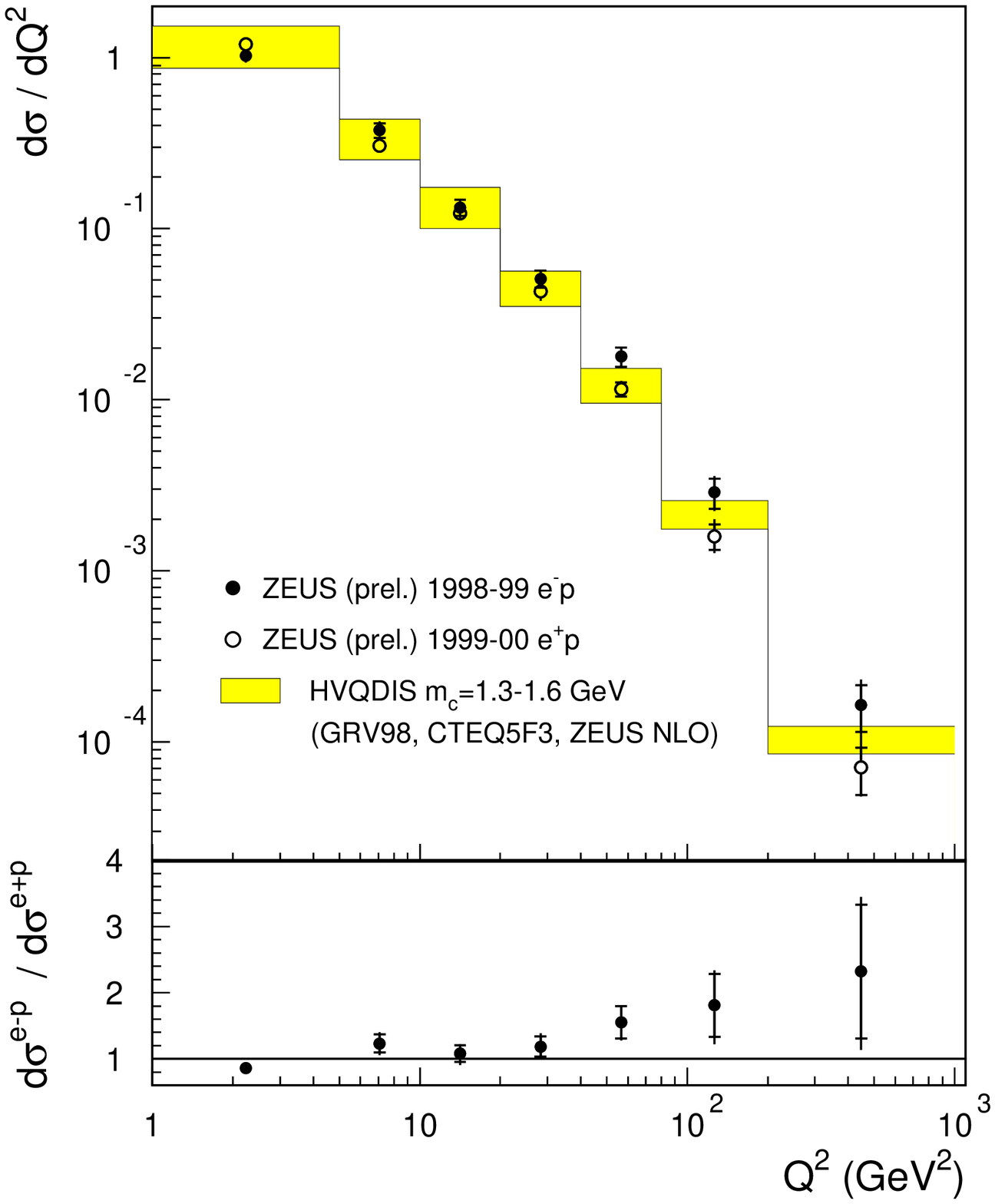,width=6.0cm}}
\end{picture}
\caption{\label{fig:eminus}
$D^{\ast\pm}$ cross section in $e^-p$ and $e^+p$ DIS
compared with NLO QCD.}}
Charm production has also been measured separately in $e^-p$ and $e^+p$ DIS 
by ZEUS~\cite{zeuseminus}. 
The data are shown in Figure~\ref{fig:eminus} as a function of $Q^2$. 
The $e^-p$ and $e^+p$ results are only barely consistent with each other; 
for $Q^2>20$~GeV$^2$, the discrepancy amounts to 3 standard deviations. 
However, both measurements are compatible with the theoretical expectation,
in which no mechanism exists to generate an asymmetry with respect 
to the lepton beam charge at such low four-momentum transfers. 

In summary, the BGF concept at NLO works well for charm in DIS, 
up to high $Q^2$.  
The HERA data reach the precision to identify regions (e.g.\ at low $x$),
where refinements are becoming necessary.

\section{Beauty}

Beauty production at HERA is suppressed 
by two orders of magnitude with respect to charm.
All HERA measurements of $b$ production 
so far rely on inclusive semi-leptonic decays,
using identified muons or electrons in dijet events. 
Two observables have been used 
to discriminate the $b$ signal from background sources. 
The high mass of the $b$ quark gives rise to large 
transverse momenta $p_T^{rel}$ of the lepton relative to 
the direction of the associated jet.
%
%\FIGURE{\unitlength1cm
%\begin{picture}(6,10.3)
%\put(-0,4.7){\epsfig{figure=dis_delta_notit.eps,width=6.5cm}}
%\put(-0.,-1.){\epsfig{figure=disptr.eps,width=6.5cm}}
%\end{picture}
%\caption{\label{fig:disptdelta}
%Muon impact parameter and $p_T^{rel}$ distributions for DIS,
%with decomposition from the likelihood fit.
%}}
%
%
%\DOUBLEFIGURE[t]{dis_delta_notit.eps,width=5.5cm}{herab.eps,width=6.cm}%
%{\label{fig:disdelta}
%Muon impact parameter distribution,
%with decomposition from the likelihood fit.}
%%{\label{fig:dispt}
%%Muon $p_T^{rel}$ distribution,
%%with decomposition from the likelihood fit.}
%{\label{fig:herab}
%Ratio of measured $b$ production cross sections at HERA 
%to theoretical expectation, as a function of $Q^2$.}
%
Using this method,
both collaborations have published 
photoproduction cross section measurements~\cite{h1openb,zeusopenb},
which are higher than NLO QCD expectations. 
\EPSFIGURE[t]{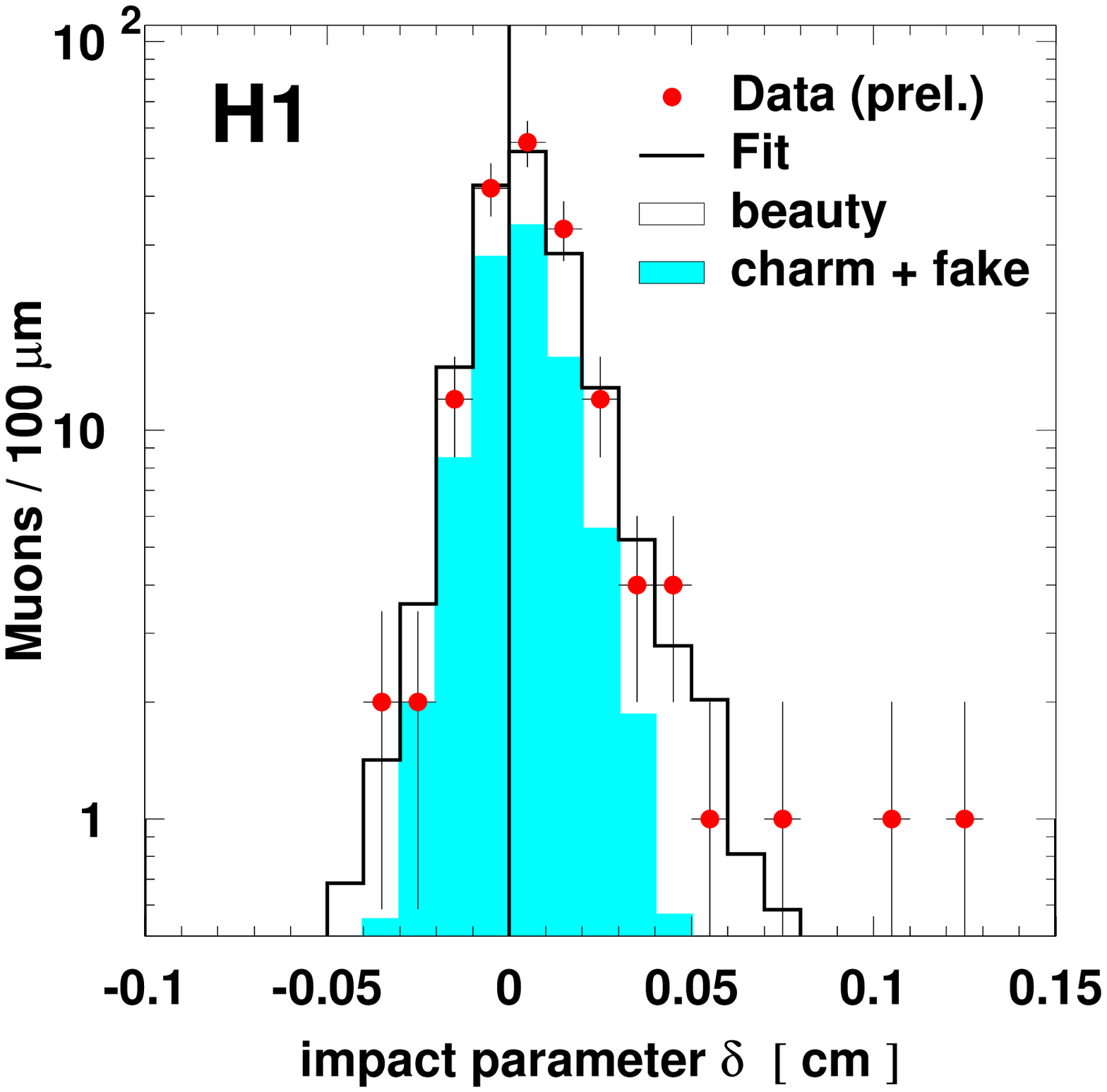,width=6.cm}
{\label{fig:disdelta}
Muon impact parameter distribution,
with decomposition from the likelihood fit.
}
More recently, 
with the precision offered by the H1 vertex detector~\cite{cst}, 
it has become possible
to observe tracks from secondary $b$ vertices
and to exploit the long lifetime as a $b$ tag, using e.g.\ the impact parameter $\delta$.
This improves the photoproduction result~\cite{h1bosaka} 
and provides a first measurement in DIS~\cite{h1bbudapest},
where resolved contributions involving the non-perturbative hadronic 
structure of the photon
are expected to be suppressed~\cite{grs}.
The DIS case is therefore complementary and theoretically simpler. 
%H1 applies the impact parameter technique to identified muons. 
The sensitivity to determine the beauty component 
is maximized by combining both variables  
in a likelihood fit to the two-dimensional 
distribution in $\delta$ and $p_T^{rel}$.
The consistency of the two observables 
has been established
with the larger statistics available in the photoproduction regime
%Requiring two jets with transverse energy above 5~GeV,
%171 muon candidates are selected from a dataset corresponding to 10.5~\pbmo.
The $\delta$ distribution for muons in dijet DIS events  
selected from a dataset corresponding to 10.5~\pbmo, 
is shown in Figure~\ref{fig:disdelta}
together with the decomposition from the two-dimensional fit,
which yields a $\bbb$ fraction of $f_b = (43\pm 8)\,\%$.
%We note that both variables are well described,
%and from both the need for a sizable $\bbb$ component is evident.
%The fit does not allow one to disentangle the background sources themselves,
%charm and fake muons, 
%with meaningful accuracy, but the $b$ fraction is only weakly sensitive 
%to the relative amount of each in the background. 
\EPSFIGURE{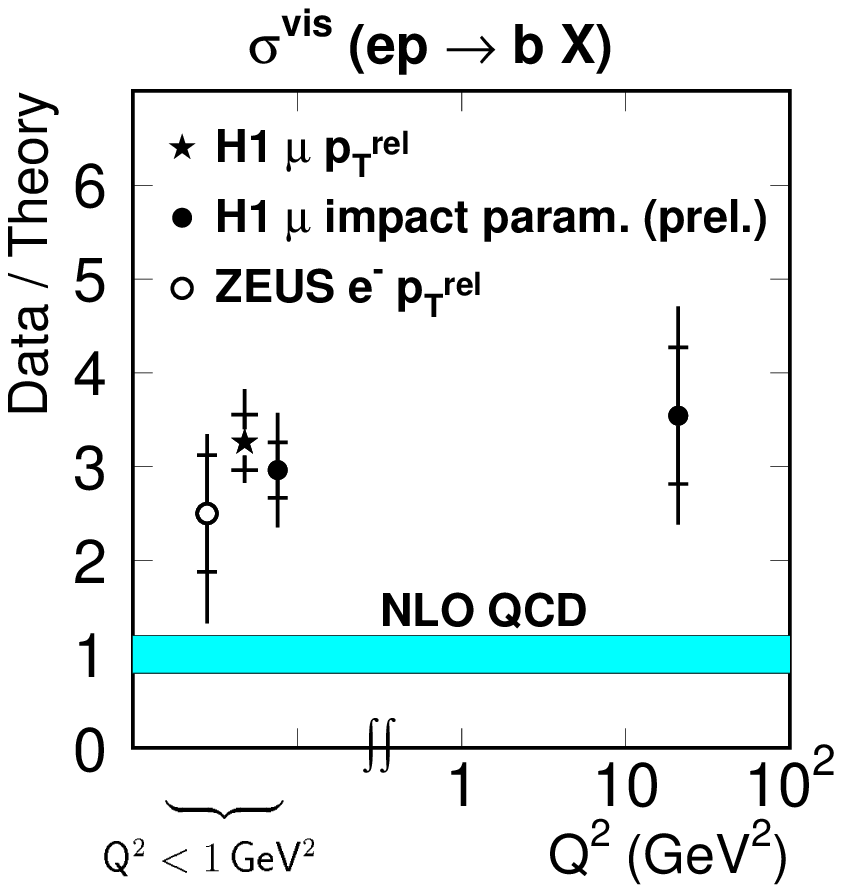,width=6.5cm}
{\label{fig:herab}
Ratio of measured $b$ production cross sections at HERA 
to theoretical expectation, as a function of $Q^2$.}
A DIS cross section of
$
\sigma_{ep\ra\bbb X\ra \mu X}^{vis} = 
%\;39\;\pm\;8\;(stat.)\; \pm 10\;(syst.)\;\;{\rm pb}\ . 
\;39\;\pm\;8\; \pm 10\;{\rm pb}\  
$
is extracted in the kinematic range 
given by $2<Q^2<100$ GeV$^2$, $0.05<y<0.7\,$, 
$p_T(\mu)>2$ GeV and 
$35^\circ<\theta(\mu)<130^\circ$.
This can be directly compared to 
NLO QCD calculations implemented in 
the HVQDIS~\cite{hvqdis} program, after folding  
the predicted $b$ hadron distributions with a decay lepton spectrum. 
The result,
$11\pm 2$~pb, 
%where the error is predominantly due 
%to the $b$ quark mass uncertainty,
is much lower than the H1 measurements. 
The data have also been compared with 
the CASCADE Monte Carlo simulation; the result
of 15~pb also falls considerably 
below the measurements.

% HERA summary
We summarize the HERA $b$ results~\cite{h1openb,zeusopenb,h1bosaka,h1bbudapest} 
as a function of $Q^2$ in Figure~\ref{fig:herab}, 
where the ratio of the measured cross sections to
theoretical expectations based on the NLO QCD calculations~\cite{hvqdis,fmnr} is displayed.
It is consistent with being independent of $Q^2$. 
The discrepancy between data and theory is similar to 
the situation observed in $\pbp$ and, 
more recently, $\gaga$ interactions~\cite{andreev-gutierrez}.
The first measurement in DIS indicates that in $ep$ collisions 
this is not a feature 
of hadron-hadron like scattering alone.

\end{document}